\documentclass[sigconf]{acmart}

\usepackage{booktabs} %

\usepackage{amsmath}
\usepackage{comment}
\usepackage{graphicx}
\usepackage{balance}
\usepackage{array}
\usepackage{amssymb}%
\usepackage{pifont}%
\usepackage{url}
\usepackage{multicol}
\usepackage{multirow}
\usepackage{array}
\usepackage{framed}
\usepackage{microtype}
\usepackage{svg}
\usepackage{tabularx}
\usepackage{xspace}
\usepackage{makecell}
\usepackage{comment}

\usepackage{balance}

\usepackage{subfig}
\usepackage[font=bf,skip=\baselineskip]{caption}

\newif\ifworkinprogress

\workinprogresstrue
\ifworkinprogress
\fi

\newcommand{\footnoteurl}[1]{\footnote{\url{#1}}}
\newcommand{\para}[1]{\smallskip\noindent\textbf{#1}}

\newcommand{\mytilde}{\raise.30ex\hbox{$\scriptstyle\mathtt{\sim}$}}

\newcommand{\ie}{\textit{i.e.}\xspace}
\newcommand{\eg}{\textit{e.g.}\xspace}
\newcommand{\cf}{\textit{cf.}\xspace}
\newcommand{\etal}{\textit{et al.}\xspace}

\newif\ifworkinprogressomments
\ifworkinprogressomments
    \newcommand{\mst}[1]{\textbf{\color{yellow}/* #1 (mst) */}}
    \newcommand{\ddi}[1]{\textbf{\color{red}/* #1 (ddi) */}}
    \newcommand{\fle}[1]{\textbf{\color{blue}/* #1 (FLO) */}}
     \newcommand{\ff}[1]{\textbf{\color{olive}/* #1 (FAB) */}}
\else
    \newcommand{\mst}[1]{}
    \newcommand{\ddi}[1]{}
    \newcommand{\fle}[1]{}
        \newcommand{\ff}[1]{}
\fi

\newcolumntype{P}[1]{>{\centering\arraybackslash}p{#1}}
\newcolumntype{M}[1]{>{\centering\arraybackslash}m{#1}}

\begin{document}
\title{Query for Architecture, Click through Military: \\ Comparing the Roles of Search and Navigation on Wikipedia}

\author{Dimitar Dimitrov}
\affiliation{%
  \institution{GESIS -- Leibniz Institute for the Social Sciences}
  \institution{\& University of Koblenz-Landau}
}
\email{dimitar.dimitrov@gesis.org}

\author{Florian Lemmerich}
\affiliation{
  \institution{RWTH Aachen University \&}
  \institution{GESIS -- Leibniz Institute for the Social Sciences}
  }
\email{florian.lemmerich@gesis.org}

\author{Fabian Fl{\"o}ck}
\affiliation{
  \institution{GESIS -- Leibniz Institute for the Social Sciences}
  }
\email{fabian.floeck@gesis.org}

\author{Markus Strohmaier}
\affiliation{
  \institution{RWTH Aachen University \&}
  \institution{GESIS -- Leibniz Institute for the Social Sciences}
  }
\email{markus.strohmaier@humtec.rwth-aachen.de}

\renewcommand{\shortauthors}{D. Dimitrov et al.}

\begin{abstract}
As one of the richest sources of encyclopedic information on the Web, Wikipedia generates an enormous amount of traffic.
In this paper, we study large-scale article access data of the English Wikipedia in order to compare articles with respect to the two main paradigms of information seeking, \ie, search by formulating a query, and navigation by following hyperlinks. 
To this end, we propose and employ two main metrics, namely (i) searchshare -- the relative amount of views an article received by search --, and (ii) resistance -- the ability of an article to relay traffic to other Wikipedia articles -- to characterize articles. We demonstrate how articles in distinct topical categories differ substantially in terms of these properties.
For example, architecture-related articles are often accessed through search and are simultaneously a ``dead end'' for traffic, whereas historical articles about military events are mainly navigated. 
We further link traffic differences to varying network, content, and editing activity features. Lastly, we measure the impact of the article properties by modeling access behavior on articles with a gradient boosting approach. The results of this paper constitute a step towards understanding human information seeking behavior on the Web.
\end{abstract}

\vspace{-2em}


\vspace{-2em}

\copyrightyear{2018} 
\acmYear{2018} 
\setcopyright{acmcopyright}
\acmConference[WebSci '18]{10th ACM Conference on Web Science}{May 27--30, 2018}{Amsterdam, Netherlands}
\acmBooktitle{WebSci '18: 10th ACM Conference on Web Science, May 27--30, 2018, Amsterdam, Netherlands}
\acmPrice{15.00}
\acmDOI{10.1145/3201064.3201092}
\acmISBN{978-1-4503-5563-6/18/05}

\maketitle

\section{Introduction}
\label{sec:intro}
Before the age of the World Wide Web~\cite{berners2000weaving}, information was predominantly consumed in a linear way, \eg, starting at the first page of a book and following the laid out narrative until the end.
With the introduction of hypertext~\cite{nelson1965complex} 
in digital environments, the way people consume information changed dramatically~\cite{leu2012new,coiro2007exploring,leu2005evaluating,mangen2008hypertext}.
While on early websites, users still predominantly visited a main page through a fixed address and were sometimes even bounded by a more directory-like navigation structure, the rise of search engines and tighter interlinking of websites have corroded the linear consumption paradigm even further. Today, users access a single website through a multitude of webpages as entry points and can usually choose from numerous paths through the available linked content at any time. In such a setting, understanding at which (kind of) pages users typically begin and end their journey on a given website, vs. which pages relay traffic internally from and to these points, provides several useful insights. On one hand, it has high practical importance since it provides the first and last contact opportunity; pages could be shaped to leverage their function as an entry point (\eg, by prioritizing improvements of navigational guidance for these pages to retain visitors), or as an exit point (\eg, by surveying visitors for their user experience before leaving, or by providing increased incentives to continue navigation).
On the other hand, knowledge about entry, relay and exit points is also closely tied to the relation of the major information seeking strategies, \ie, search and navigation:
the first page visited in a session on a website is frequently reached via search engine results, after a query formulation, while navigation has been often used when the exact information need cannot be easily expressed in words~\cite{furnas1997effective,furnas1987vocabulary}. 
Understanding under which circumstances search or navigation dominate the users' information seeking behavior can help in developing an agenda for improving the web content in order to optimize visitor rates and retention.

\para{Scope and research questions.} Information consumption on the Web has been of special interest to researchers since the Web's earliest days~\cite{kumar2010characterization,kumar2009characterization}. 
While both search ~\cite{waller2011search,mcmahon2017substantial, spoerri2007popular} and navigation ~\cite{dimitrov2017makes,gilderslave2017Inspiration,lehmann2014reader,lamprecht2016evaluating,lamprecht2017structure} have been investigated thoroughly in related work, they were mostly looked at separately. Consequently, so far little is known about which parts and content types of a specific website (inter)act in which structural roles, begetting different information access patterns. 

In this work, we analyze how these patterns manifest on the online encyclopedia Wikipedia. With more than 5 million articles, Wikipedia is one of the primary information sources for many Web users and through its openly available pageview data provides an essential use case for studying information seeking behavior, as made apparent by numerous studies~\cite{dimitrov2016visual,lamprecht2017structure,paranjape2016improving}.
Yet, there is a lack of understanding how search and navigation as the two major information access forms \textit{in combination} shape the traffic of large-scale hypertext environments, such as the world's largest online encyclopedia. %
To this end, we are interested in answering the following research questions: 
(i) How do search and navigation interplay to shape the article traffic on Wikipedia? Given an article, we want to know how its acting as a search entry point is related to (not) relaying navigation traffic into Wikipedia, and vice versa. This also addresses the issue of how search and navigation contribute to the article's popularity. Beyond these characteristics of the system in general, we also examine which specific properties of articles influence their roles in the search-vs-navigation ecosystem. We hence ask: (ii) Which article features (\ie, topic, network, content and edit features) are indicative for specific information access behavior? 

\para{Materials, approach and methods.} Building our analysis on large-scale, openly available log data for the English edition of Wikipedia, we propose two metrics capturing individual traffic behavior on articles, \ie, (i) searchshare -- the amount of views an article received by search --, and (ii) resistance -- the ability of an article to channel traffic into and through Wikipedia (\cf Section~\ref{sec:traffic_def}).

We use searchshare and resistance to first explore the relation between search and navigation and their effect on the popularity of articles independent of their content (\cf Section~\ref{sec:gab}).
Depending on these two measures, we assign articles to four groups describing the role they assume for attracting and retaining visitors. 
Subsequently, we characterize the influence of several article attributes, including the general topical domain, edit activity and content structure on the preferred information access form (\cf Section~\ref{sec:ch}). Finally, we fit a gradient boosting model to determine the impact of these article features on the preferred user access behavior (\cf Section~\ref{sec:prediction_task}).

\para{Contributions and findings.} 
Our contributions are the following: (i) Regarding the general (collective) access behavior on Wikipedia, we provide empirical evidence that for the most viewed articles search dominates navigation in the number of articles accessed and received views. For the tail of the view distributions, navigation appears to become more and more important. (ii) We link article properties, \ie, position in the Wikipedia network, number of article revisions, and topic to preferred access behavior, \ie, search or navigation. Finally, (iii) we quantify the strength of the relationship between article properties and preferred access behavior.

Our analysis suggests that (i) while search and navigation are used to access and explore different articles, both types of information access are crucial for Wikipedia, and (ii) that exit points of navigation sessions are located at the periphery of the link network, whereas entry points are located at the core. 
(iii) Edit activity is strongly related with the ability of an article to relay traffic, and thus with the preferred access behavior.

Our results may have a variety of applications, \eg, improving and maintaining the visual appearance and hyperlink structure of articles, identifying articles exhibiting changes in access behavior patterns due to vandalism or other online misbehavior. 
We consider our analysis as an initial step to better understand how search and navigation interplay to shape the user access behavior on platforms like Wikipedia and on websites in general.

\section{Transition Data and Definitions}
\label{sec:traffic_def}
Below, we give an overview of the used dataset capturing the traffic on Wikipedia articles and define \textit{searchshare} and \textit{resistance} as our main metrics for describing the individual article traffic behavior.

\subsection{Transition Data}
\label{sec:data}
For studying the access behavior on Wikipedia articles, we use the clickstream dataset  published by the Wikimedia Foundation~\cite{clickstream}.
The used dataset contains the transition counts between webpages and Wikipedia articles in form of (\textit{referrer, resource}) pairs extracted from the server logs for August, 2016, and is limited to pairs that occur at least 10 times. The referrer pages are either external (\eg, search engines, social media), internal (other Wikipedia pages), or missing (\eg, if the article is accessed directly using the browser address bar). The navigation targets are purely internal pages.\footnote{Leaving a Wikipedia page is treated as the end of the visit in the logs, whether by clicking on an external link or closing the page.} 
Since we are interested in contrasting Wikipedia article access from search engines and navigation (see also our discussion in Section \ref{sec:discussion}),  we focus our analyses only on those articles in the clickstream dataset that have received views through search or internal navigation, setting aside remaining view sources (mostly "no referrer").
Accordingly, we define \textit{total views} of an article as the sum of all page accesses by either search or navigation.

The resulting dataset consists of 2,830,709 articles accessed through search 2,805,238,298 times and 14,405,839 transitions originating from 1,370,456 articles and accounting for 1,251,341,103 views of 2,149,104 target articles. In total, the dataset consists of 3,104,702 articles viewed 4,056,579,401 times, with a ratio of 69\% stemming from search and 31\% from internal navigation -- in line with previous reports on the clickstream data~\cite{dimitrov2017makes,lamprecht2016evaluating}.
\ddi{854473 articles are accessed exclusively through search, those are the nodes in the other component}

\subsection{Definitions}
\label{sec:entry_exit}
To achieve a fundamental understanding of the parts that search and navigation each play for the distribution of views in Wikipedia, we take a look at %
the functional roles articles can assume for the overall traffic flow in respect to their \textit{searchshare} and \textit{resistance}.

\para{Searchshare.}
A high \emph{searchshare} value indicates that search is the predominant paradigm of accessing an article, and thus that the article acts as an \textit{entry point} for a site visit. In contrast, articles with a low value receive most of their views from users visiting them by means of  navigation.  
The \textit{searchshare} metric is defined as
\begin{align}
searchshare (a) = \frac{in_{se}(a)}{in_{se}(a) + in_{nav}(a)} 
\end{align}
where $in_{se}(a)$ is the number of pageviews an article $a$ received directly from  search engine referrers, and $in_{nav}(a)$ is the number of views from navigation as recorded in the Wikipedia clickstream.

\para{Resistance.}
A low \textit{resistance} value signals that an article forwards most of its received traffic to other articles within Wikipedia, hence does not block the flow of incoming traffic onward. A high value in turn indicates that an article acts as an  
\emph{exit point}. Thus, it rarely relays users to other Wikipedia articles. These articles are traffic sinks in the Wikipedia information network.
We define the resistance metric as
\begin{align}
resistance (a) = 1 - \frac{out_{nav}(a)}{in_{se}(a) + in_{nav}(a)} 
\end{align}
where $out_{nav}(a)$ is the number of pageviews that had article $a$ as a referrer.
Additionally, we restrict the values to be in the interval [0,1]. This is necessary since a small number of articles generates more out-going traffic than they receive pageviews, \eg, due to a user opening several links in a new tab each.

\begin{figure}[t]
\centering
\subfloat[Search and navigation vs. total]{\includegraphics[width=0.245\textwidth]{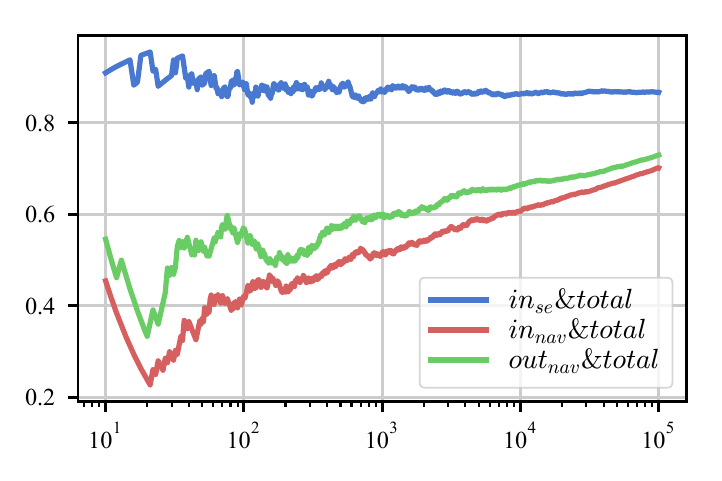}\label{fig:overlap_total}}
\subfloat[Search vs. navigation]{\includegraphics[width=0.245\textwidth]{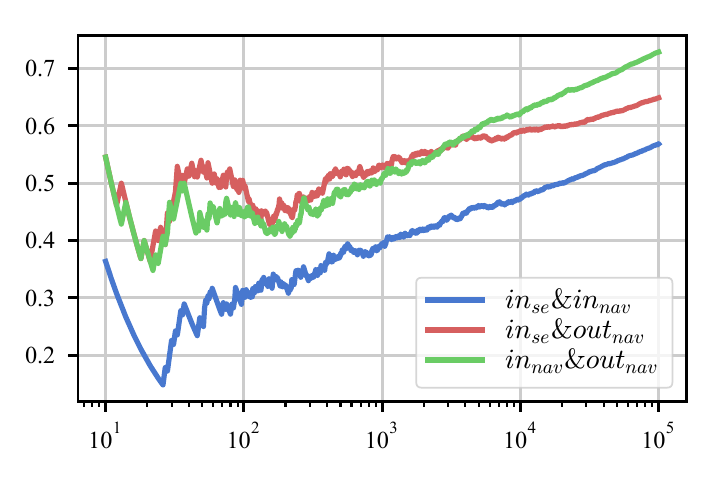}\label{fig:overlap_mixed}}
\caption{Ranking overlap. Four rankings are shown, according to the total number of pageviews (\textit{total}), the number of pageviews coming from search ($in_{se}$) as well as in- ($in_{nav}$) and out-navigation ($out_{nav}$).  
The y-axis indicates overlaps between pairs of rankings, considering the top-k articles of each ranking as marked on the x-axis (log-scaled, top articles on the left).
As a result, the overall ranking of total pageviews shows a very high overlap with the incoming search ranking. The top pages by search and navigation differ substantially. Notably, being a distribution point of traffic (high $out_{nav}$, \cf (b)) is correlated most to receiving search, but only for top $out_{nav}$ articles, with lower ranks being supplied with traffic predominantly through $in_{nav}$.}
\label{fig:overlap}
\vspace{-1em}
\end{figure}

\section{General Access Behavior}
\label{sec:gab}
In this section, 
we investigate how exogenous and endogenous traffic contribute to article popularity on Wikipedia, and we study the distribution of traffic features. 
We provide a first %
overview of the general access behavior on Wikipedia regarding search and navigation, aided by a division of articles into four groups with respect to searchshare and resistance; in Section \ref{sec:ch}, we will subsequently take a deeper look at dissimilarities between different types of articles. 

\para{Search and navigation in relation to total views.}
As can be expected from related research on Wikipedia and similar online platforms, the distribution of pageviews over articles is long-tailed with a heavy skew towards the head (80\% views generated by the top-visited 5.2\% of all articles).
To better investigate the relationship between search and (incoming and outgoing) navigation on the articles popularity, we calculate the cumulative overlap (intersection) of the descendingly ranked articles at each rank \textit{k}, divided by \textit{k}; this is an adaptation of the Rank Biased Overlap\footnote{Rank Biased Overlap~\cite{webber2010similarity} is a common metric for similarity between rankings using cumulative set overlap in cases where the two lists do not necessarily share the same elements (as is the case here). Top-weighting as can be specified for RBO is neither suited nor necessary for the distinction of different \textit{k} that we aim for here.} measure.

\begin{figure}[t!]
\centering
\subfloat[Number of articles]{\includegraphics[width=0.245\textwidth]{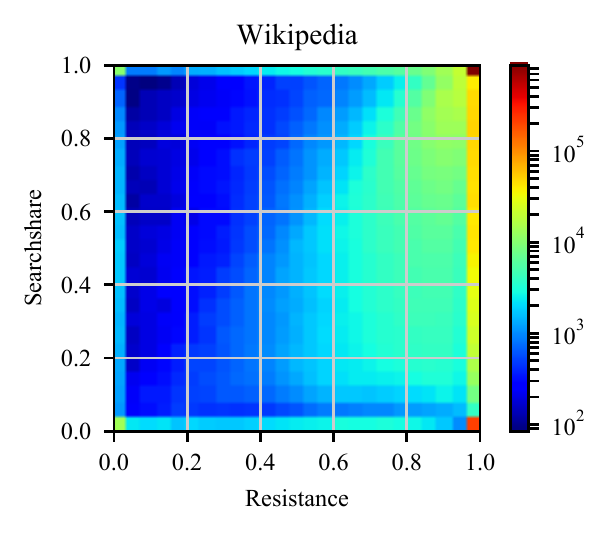}\label{fig:heatmap_all}}
\subfloat[Total views]{\includegraphics[width=0.245\textwidth]{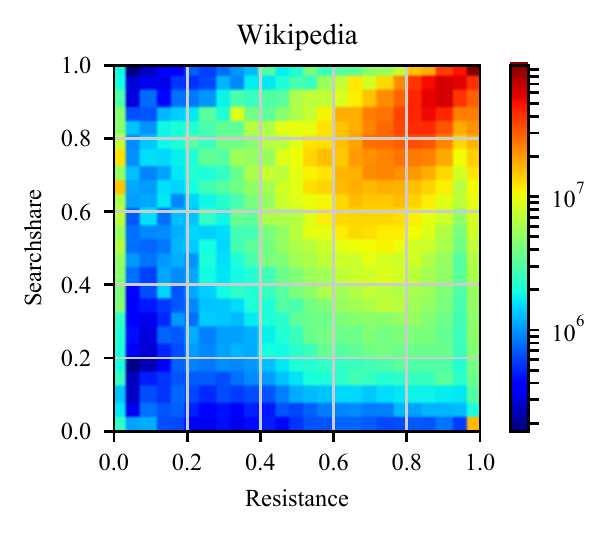}\label{fig:heatmap_pageviews_all}}
\caption{Articles and article views by access behavior. For a given searchshare (y-axis) and resistance (x-axis), the figure shows (a) the number of articles and (b) the sum of their views in each heatmap square bin. Warm colors denote high values, using a logarithmic scale. We observe that search  dominates navigation in terms of number of accessed articles (note the single top data bin in (a)) and that a substantial amount of articles exhibits high resistance values. When focusing on views, we see a more spread-out pattern, evidencing that a relatively small amount of articles attracts a substantial amount of search views and channels them onward to other articles (upper left side of (b)), corresponding to the \textit{search-relay} group (\cf Table~\ref{tab:quadrants}).}
\label{fig:general_heatmap}
\vspace{-1em}
\end{figure}

\begin{figure*}[t]
\centering
\subfloat[Searchshare]{\includegraphics[width=0.25\textwidth]{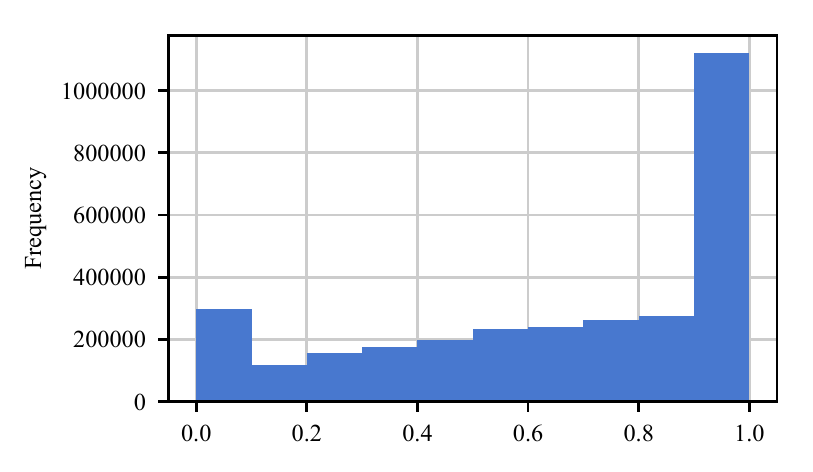}\label{fig:hist_a}}
\subfloat[Resistance]{\includegraphics[width=0.25\textwidth]{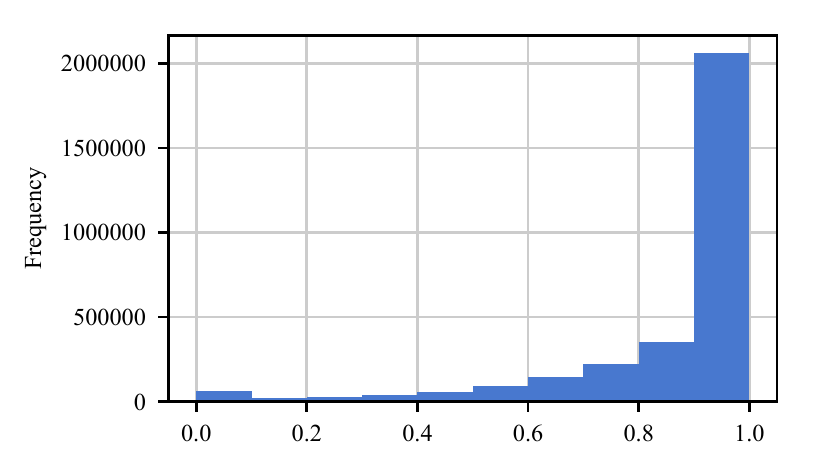}\label{fig:hist_b}}
\subfloat[Searchshare-weighted]{\includegraphics[width=0.23\textwidth]{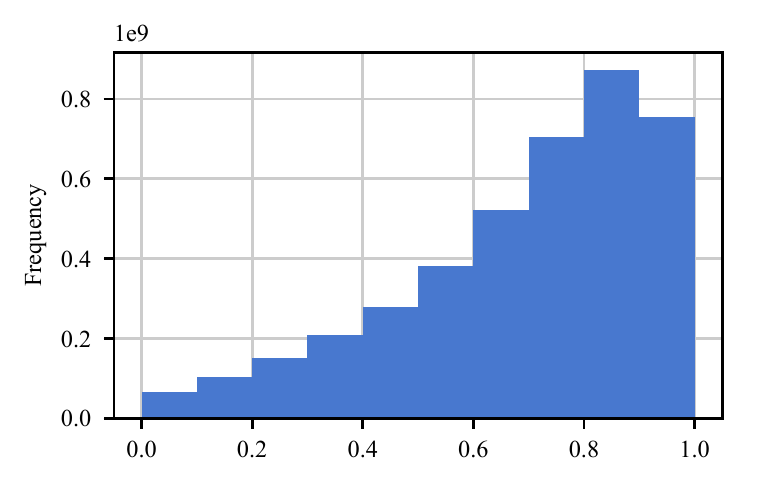}\label{fig:hist_w_a}}
\subfloat[Resistance-weighted]{\includegraphics[width=0.23\textwidth]{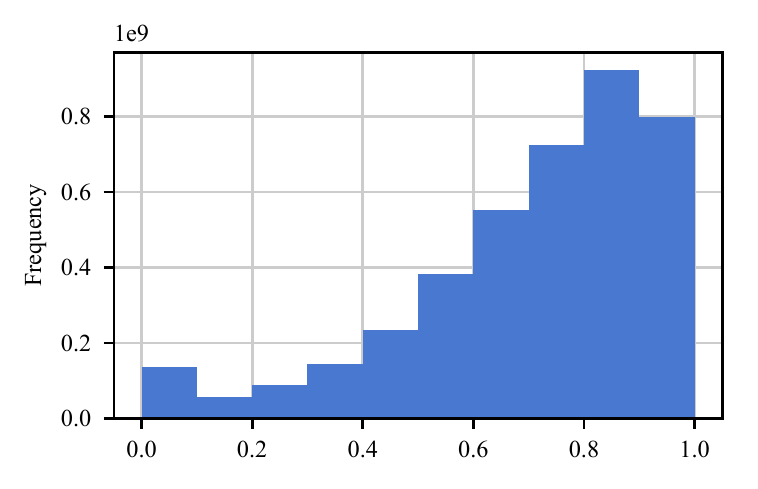}\label{fig:hist_w_b}}
\vspace{-1em}
\caption{Traffic feature distributions. Figures (a) and (b) show an unweighted histogram of searchshare and resistance, while (c) and (d) respectively weight  articles by their pageview counts. Most articles have a very high value for searchshare and resistance. However, extreme values close to $1.0$ in (a), (b) stem mostly from rarely visited pages.%
}
\label{fig:hist}
\vspace{-1em}
\end{figure*}

Figure~\ref{fig:overlap}\subref{fig:overlap_total} shows that the top \textit{k} articles  ordered by search traffic (top-k-search) are highly overlapping with the top articles by total views (top-k-total) at any \textit{k}, underscoring the general importance of search as a driver of incoming views. 
In-navigation, in contrast, is not a deciding factor to belong to the top most visited pages, but sees an extreme increase in the influence on overall views for articles up to top-k-total around 8000, at which point the increase continues, but levels off. Apparently, while search is the overall main driver for traffic, in-navigation rapidly becomes a more central source of traffic beyond the extremely popular articles. 
Turning to navigation passed on \textit{from} articles to other articles, we can glean from Figure~\ref{fig:overlap}\subref{fig:overlap_total} that (i) while the very top of viewed articles contribute little in relation to their accumulated views to the internal traffic flow of Wikipedia (low overlap for $out_{nav}\&total$), we (ii) see a rapid and constant drop in the amount of traffic ``dying'' at a given page with increasing top-k-total.

Further, while it is not surprising that the outgoing traffic accumulates generally in line with the overall received views, up until around top-k-total 1,500,000 it is generated at a rate \textit{surpassing} the relative increase of total views, with the highest ranks of top-k-total contributing comparably little to it, just as to in-navigation. 
These observations are in line with Figure~\ref{fig:overlap}\subref{fig:overlap_mixed}, where we see that a higher rank in receiving navigation - rather than from search - is more strongly correlated with distributing views to other articles for the largest portion of pages, after top-k-total 3000; up until that point, the largest share of channeled traffic stems from search views. 
As bottom line, we see a pattern that points to a small number of pages at the extreme top of the pageview counts that are mostly searched, but \textit{in relation to their popularity} rather isolated in terms of navigation; 
with in- and out-navigation similarly gaining notably in correlation with overall views for lower top-k-total ranks.

\begin{table}[b]
\centering
\small
\def\arraystretch{0.9}
\setlength\tabcolsep{0.4em}
\vspace{-1em}
\caption{Article group sizes and views. For each group, the table shows the percentage of articles and their received views. The majority of the articles are less visited and act as exit points of user session, whereas only popular articles are able to further relay traffic.}
\begin{tabular}{|c|c|c|c|c|c|}
\toprule
            & search-exit & search-relay & nav.-relay      & nav.-exit & total \\
\midrule
articles    &    43\%    &   9\%       &     21\%  &  27\% & 100\% \\
views       &    17\%    &   37\%       &     39\%  &  7\% & 100\% \\
\bottomrule
\end{tabular}
\label{tab:quadrants}
\vspace{-1em}
\end{table}

\para{Traffic feature distributions.}
Figure~\ref{fig:hist} depicts the system-wide distribution of searchshare and resistance.
Pages are generally much more searched than navigated to (searchshare median = 0.74, mean = 0.66) as seen in Figure~\ref{fig:hist}\subref{fig:hist_a}. It is also apparent from Figure~\ref{fig:hist}\subref{fig:hist_b} that most articles do not tend to forward much of their received traffic internally, with the median for resistance for all articles lying at 1.0 and the mean at 0.88. This general tendency prevails when these scores are weighted by their received views (Figures~\ref{fig:hist}\subref{fig:hist_w_a} and~\ref{fig:hist}\subref{fig:hist_w_b}), but a notably less skewed distribution emerges, implying that -- even when accounting for regression-to-the-mean effects -- a majority of views is acquired via search and that a majority of views hits rather high-resistance targets.

\para{Relation between searchshare and resistance.}
We observe a light positive correlation  (pearson = 0.26, spearman = 0.33) indicating that the more likely an article is used to start a session, the more likely it is also to be the last article accessed in a session. Figure~\ref{fig:general_heatmap} depicts this association for all articles in our dataset. 

To explore this relation further, we assign each article to one of four groups, determined by the \textit{mean} of both searchshare and resistance as the  thresholds.\footnote{A delimitation by median yields groups with the sole resistance value 1.0 and was therefore not used. Cut-offs at 0.5 would have created extremely unbalanced groups.} %
We label each group according to its traffic behavior, \ie, (i) \textit{search-relay} articles that are often searched while simultaneously contributing to further navigation (above-mean searchshare, below-mean resistance); (ii) \textit{search-exit} articles with above-average searchshare that are often accessed from search but do not lead to users navigating further (above-mean searchshare, above-mean resistance); (iii) \textit{navigation-exit} articles that receive their traffic mostly from navigation but cannot channel traffic to other pages (below-mean searchshare, above-mean resistance); (iv) \textit{navigation-relay} articles that are mainly accessed from within Wikipedia and able to pass traffic on internally (below-mean searchshare, below-mean resistance).
Table \ref{tab:quadrants} reports the share of articles and views pertaining to each group. We observe that a small group of  highly visited articles is able to inject considerable amounts of traffic (search-relay) into Wikipedia while about a fifth of the articles' role is mainly to channel traffic internally (nav.-relay). On the other hand, exit points receive less views while covering a much bigger portion of Wikipedia articles. Overall, these observations are in line with Figure~\ref{fig:hist}.

\para{Summary.} Our analysis shows that search dominates navigation with respect to the number of articles accessed and visit frequency. However, the less viewed an article is, the more significant navigation becomes as an information access form. Further, only popular articles are able to relay traffic while the majority of the articles acts as exit points for user search and navigation sessions.

\section{Characterizing Access Behavior}
\label{sec:ch}
In the previous section, we analyzed the general Wikipedia information access behavior, setting aside individual page attributes. However, Wikipedia articles have different properties that may influence the way they are retrieved (\cf Section~\ref{sec:article_features}). To this end, we analyze the general Wikipedia access behavior dependent on the article network (\cf Section~\ref{sec:network}), and content and edit properties (\cf Section~\ref{sec:content}). Subsequently, we highlight differences between general access behavior on Wikipedia and on Wikipedia topics dominated by search and navigation, respectively (\cf Section~\ref{sec:initail_topic}).

\subsection{Wikipedia Article Data and Features}
\label{sec:article_features}
To study the influence of the content on the preferred access behavior, we focus on a snapshot of all Wikipedia articles contained in the main namespace of the English language version from August, 2016\footnote{\url{https://archive.org/details/enwiki-20160801}}. We obtained the articles using the Wikipedia API\footnote{\url{https://www.mediawiki.org/wiki/API:Main_page}}. The collected article data represent the %
HTML version of each article on which the transitions data used to study the Wikipedia traffic has been generated (\cf Section~\ref{sec:data}). By parsing and rendering the HTML version of the articles, we are able to extract article features capturing aspects related to the content of the articles. The dataset contains roughly 5 million articles connected by 391 million links.

For these Wikipedia articles, we determine a wide variety of features describing their characteristics.
We categorize these features into three different groups, \ie, (i) network properties, (ii) content and edit properties and (iii) article topics.
The network features consist of \textit{in-, out-} and \textit{total degree} of the article in the Wikipedia link network as well as the k-core value for this network as a typical centrality measure. 
Regarding the content and edit properties, we calculated for each article the \textit{number of sections}, the \emph{number of figures} and the \emph{number of lists} contained in the article. These features capture visual appearance of the article, whereas the \textit{number of revisions and editors} represent content production process. We also consider the article \textit{age} measured in years to account for differences between mature and young articles. To account for the amount of information provided in an article, we calculate its \textit{size in kilobytes}. The features capturing the content production process are extracted from the TokTrack dataset~\cite{ICWSM1715689} and consider the period between article creation and the end of August 2016. As the Wikipedia article categories are often too specific\footnote{I.e., very specific categories of articles are not linked to the relevant super-category; in other cases, two conflicting categories are linked or fitting categories are missing completely. }, we fit a Latent Dirichlet Allocation (LDA)~\cite{blei2003latent} model on article texts using Gensim~\cite{rehurek_lrec} bag of words article vectors with removed stop words. To allow for manual interpretation of the topics, we fit a model for 20 topics. Subsequently, we asked five independent researchers to provide topic labels based on the top words and Wikipedia articles for each topic and summarized their labels. Section \ref{sec:initail_topic} describes the extracted topics. The following analyses are based on a random sample of 50000 articles.

\begin{figure}[t]
\centering
\subfloat[k-core vs. searchshare]{\includegraphics[width=0.245\textwidth]{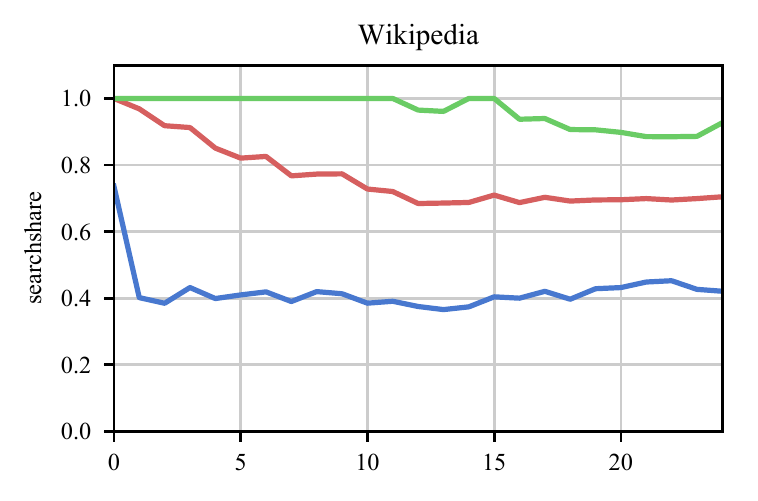}\label{fig:n_a}}
\subfloat[k-core vs. resistance]{\includegraphics[width=0.245\textwidth]{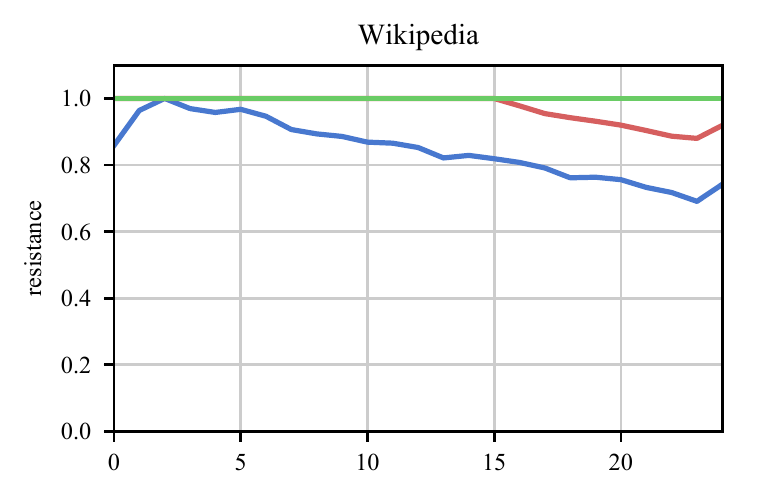}\label{fig:n_b}}
\caption{Network position. The figure shows the first (blue), second (red), and third (green) quartile of searchshare (a) and resistance (b) as function of the article position in the network indicated by its k-core. K-core values are divided into 25 bins. The access behavior on articles is influenced by their position in the network. The more central an article, the lower its searchshare and resistance -- i.e., the more traffic it relays through the network.}
\label{fig:network}
\vspace{-1em}
\end{figure}

\subsection{Network Features}
\label{sec:network}
\begin{table}[b]
\centering
\small
\def\arraystretch{0.9}
\setlength\tabcolsep{0.4em}
\vspace{-1em}
\caption{Network features.  For each network feature, the table shows the \textit{median} feature values of the articles in the respective group. The article network properties influence the preferred access behavior. Nav.-relay articles act as intersections for the traffic as they occupy central network positions and provide lots of in- and outgoing links. Search-relay articles are similarly well-connected, which is important for injecting traffic into Wikipedia. Exit points (search-exit and nav.-exit articles) lack connectivity and are unable to channel external and internal traffic, respectively.}
\begin{tabular}{|c|c|c|c|c|c|}
\toprule
   $M$         & search-exit & search-relay & nav.-relay      & nav.-exit & overall\\
\midrule
in-degree   &    14      &   38       &     54      &  18   &  22 \\
out-degree  &    33      &   56       &     71      &  35   &  41 \\
degree      &    51      &   105      &     131     &  57   &  69 \\
k-core      &    44      &   76       &     95      &  49   &  57 \\
\bottomrule
\end{tabular}
\label{tab:quadrants_network}
\end{table}

To understand the role of the network features, we compute the median of the features for each of the four article groups \emph{search-exit}, \emph{search-relay}, \emph{nav.-relay}, and \emph{nav.-exit} (\cf Section~\ref{sec:gab}).
The results are shown in Table~\ref{tab:quadrants_network}.
We can observe that articles with below-average searchshare and resistance (article group \emph{nav.-relay}) have higher median values across all network features, \ie, they are located more in the center of the network and consistently have more incoming and outgoing links. Although search-relay articles are not as well connected as nav.-relay, their relatively central position in the network and high number of outgoing connections is important in order to inject traffic into Wikipedia.
By contrast, articles that are often used as exit points (\emph{search-exit} and \emph{nav.-exit}) are located more in the periphery of the network (low k-core value), are less often linked to, and contain less out-links themselves, which eventually results in higher resistance values, signifying the termination of user sessions. 

For further analysis, we sort the articles according to their k-core value and discretize them into 25 equally-sized bins. For each bin, we compute the quartiles for searchshare and resistance, as seen in Figure~\ref{fig:network}.
Looking at the median (center red line), we find that  for articles with increasing k-core values the searchshare indeed decreases (\cf Figure~\ref{fig:network}\subref{fig:n_a}). However, this effect stops at around 50\% of the dataset, \ie, for half of the articles, which are located in high k-core network layers, the searchshare is mostly independent from the exact centrality.
Regarding the resistance, there exists a substantial amount of nodes with a resistance of 1.0 for all k-core values, \cf the green line indicating the upper quartile. However, for the more central nodes, an increased number of pages have a significantly lower resistance (\cf Figure~\ref{fig:network}\subref{fig:n_b}).

\subsection{Content and Edit Features}
Next, we characterize the article groups in terms of the article content and edit history which account for the content presentation and content production process.
Table~\ref{tab:quadrants_edit} reports the median values of these features in the four article groups.
We can observe that the content features (number of tables, number of sections, size of the article) are modestly increased for relay articles, \ie, articles that contain more content tend to be less often exit points of navigation sessions.
By contrast, the revision history plays a more important role: we can see that articles in the \emph{search-relay} group have (as a median) more than twice the number of editors and revisions compared to exit articles, and tend also to be somewhat older. Articles in the group \emph{nav.-relay} show similar, but slightly lower values with the same tendency.
The median feature values for both ``exit'' article groups are very similar and show slightly lower editor and revision numbers.
Overall, content and edit features provide strong indicators for articles relaying traffic (as opposed to being exit points), but only weak indicators for being accessed by search or by navigation.

\begin{figure}[t]
\centering
\subfloat[revisions vs. searchshare]{\includegraphics[width=0.245\textwidth]{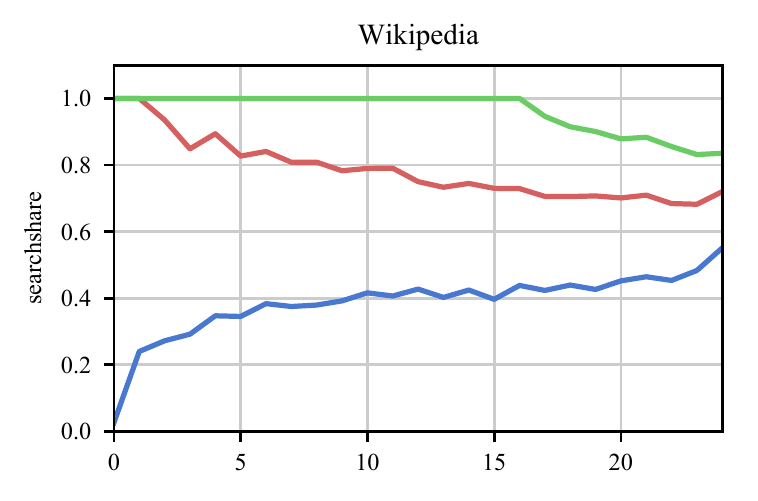}\label{fig:e_a}}
\subfloat[revisions vs. resistance]{\includegraphics[width=0.245\textwidth]{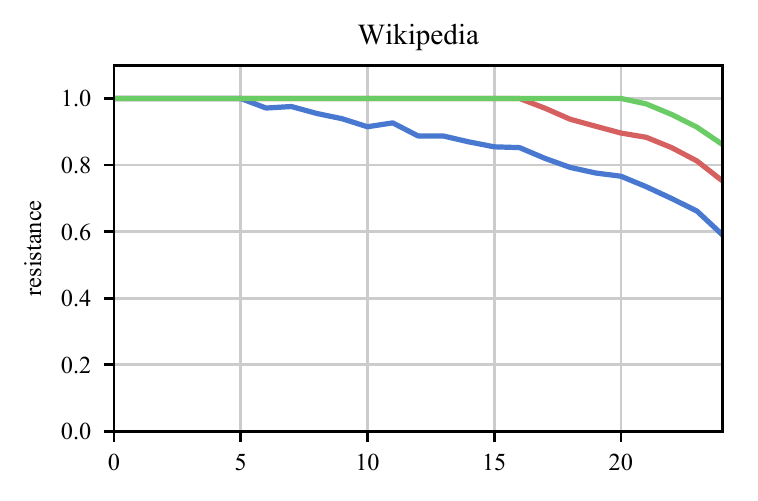}\label{fig:e_b}}
\caption{Edit activity. The figure shows the first (blue), second (red), and third (green) quartile of searchshare (a) and resistance (b) as function of the article editors' activity indicated by the number of revisions. Revisions values are divided into 25 bins. Except for the most edited articles, high edit activity has a negative effect on the resistance, which on the other hand has a positive effect on navigation indicated by the lower searchshare.}
\label{fig:edit}
\vspace{-1em}
\end{figure}

\label{sec:content}
\begin{table}[b]
\centering
\small
\def\arraystretch{0.9}
\setlength\tabcolsep{0.4em}
\vspace{-1em}
\caption{Content and edit features. For each content and edit feature, the table shows the median feature values of the articles in the respective group. The content production process influences the access behavior as search- and nav.-exit points have low edit activity, and offer less content. On the other hand, relay articles are more frequently edited, and congruently, are generally more extensive.}
\begin{tabular}{|c|c|c|c|c|c|}
\toprule
    $M$        & search-exit & search-relay & nav.-relay      & nav.-exit & overall \\
\midrule
editors     &    21      &   52       &     46      &  21  &  25   \\
revisions   &    38      &   97       &     86      &  37  &   46  \\
sections    &    6       &   7        &     7       &  4   &    6  \\
tables      &    3       &   3        &     4       &  3   &   4   \\
age         &    9       &   11       &     10      &  8   &    9  \\
size        &    41      &   50       &     54      &  41  &   44  \\
\bottomrule
\end{tabular}
\label{tab:quadrants_edit}
\end{table}

We will have a more detailed look at an exemplary edit feature, \ie, the number of revisions. Analogously to above (network features), we assign the articles to one of 25 bins according to their revision count, compute the distribution of searchshare and resistance for each bin, and plot the quartiles.
The results are shown in Figure~\ref{fig:edit}.
We can see that the median searchshare continuously decreases with increasing number of revisions. 
The effect is in particular significant for very low number of revisions (\cf Figure~\ref{fig:edit}\subref{fig:e_a}).
Additionally, the spread of the distribution -- measured by the interquartile range (IQR) -- also substantially decreases the more revisions an article has. 
This can likely be explained by \emph{regression to the mean} since articles with less revisions receive overall less views, making more extreme searchshare values more likely.
With regard to the resistance, we can observe that specifically high number of revisions correlate with a lower resistance scores (\cf Figure~\ref{fig:edit}\subref{fig:e_b}).
The number of editors, and the age of an article is highly correlated with the number of revisions and reveal a very similar behavior with respect to searchshare and resistance.

\begin{figure}[t]
\centering
\subfloat[\textbf{searchshare}]{\includegraphics[width=0.245\textwidth]{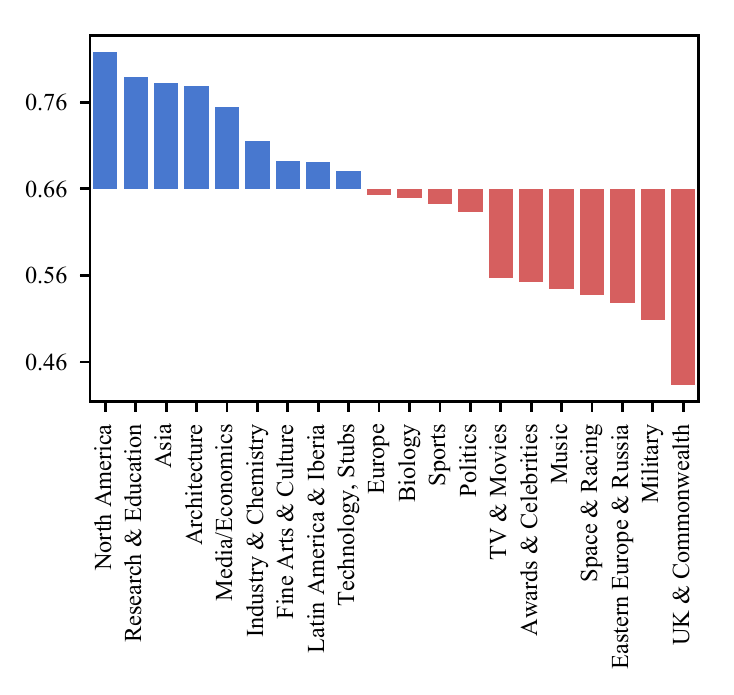}\label{fig:t_a}}
\subfloat[\textbf{resistance}]{\includegraphics[width=0.245\textwidth]{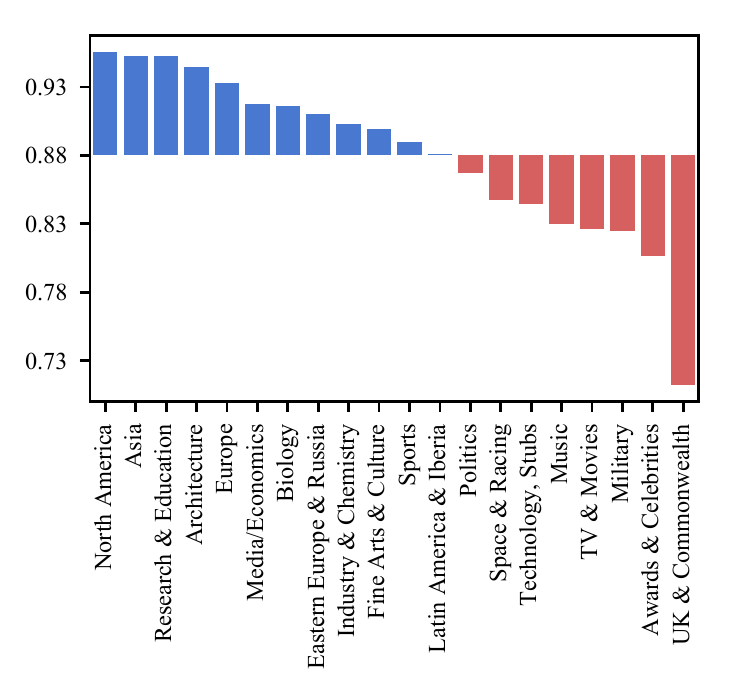}\label{fig:t_b}}
\caption{Access behavior for all topics. Topics are ordered from highest (left) to lowest (right) for searchshare (a) and resistance (b).  Values over (blue) and below (red) the respective mean value are colored respectively. There are pronounced differences in the dominant access behavior on different Wikipedia topics.}
\vspace{-1em}
\label{fig:topic_views}
\end{figure}

\begin{table}[b]
\centering
\small
\setlength\tabcolsep{0.4em}
\def\arraystretch{0.9}
\vspace{-1em}
\caption{Topic statistics. The table shows the percentage of articles and views for each topic. Additionally, it reports the median age in years, number of editors and revisions, and the size of the articles in kB. 
Not surprisingly, popular articles are generally longer in terms of text, edited more and by higher number of editors, and relatively old.}
\label{tab:topics}
\begin{tabular}{|c|c|c|c|c|c|c|c|}
\toprule
                   topic &  \makecell{\% \\articles} & \makecell{\% \\views} &  \makecell{$M$\\age} & \makecell{$M$\\editors} &   \makecell{$M$\\revisions} &  \makecell{$M$\\size} \\
\midrule
       Technology, Stubs &        19.3 &         7 &  7 &     20 &   36 &  38 \\
            Architecture &        12.4 &         5 &  8 &     31 &   61 &  56 \\
                  Sports &        12.0 &         8 &  7 &     37 &   86 &  68 \\
                Politics &        8.1 &         8 &  8 &     47 &  103 &  60 \\
             TV\&Movies &       7.5 &         32 &  8 &     96 &  197 &  55 \\
     Fine Arts\&Culture &       7.2 &         6 &  8 &     49 &  100 &  49 \\
                 Biology &        7.0 &         6 &  7 &     29 &   57 &  46 \\
                   Music &        6.9 &         9 &  8 &     64 &  136 &  53 \\
    Research\&Education &       4.8 &         2 &  7 &     40 &   87 &  47 \\
         Media/Economics &        3.3 &         4 &  8 &     54 &  115 &  49 \\
                Military &        3.1 &         4 &  8 &     47 &  105 &  65 \\
    Industry\&Chemistry &       3.0 &         6 &  9 &     66 &  126 &  55 \\
           North America &        1.2 &         0 &  9 &     23 &   39 &  52 \\
          Space\&Racing &       1.2 &         2 &  8 &     52 &  114 &  73 \\
                  Europe &        0.9 &         0.0 &  7 &     19 &   36 &  52 \\
                    Asia &        0.7 &         0.0 &  5 &      8 &   14 &  60 \\
  Latin America\&Iberia &       0.5 &         0.0 &  7 &     20 &   33 &  58 \\
       UK\&Commonwealth &       0.5 &         0.0 &  7 &     19 &   40 &  43 \\
 Eastern Europe\&Russia &       0.4 &         0.0 &  7 &     15 &   24 &  49 \\
    Awards\&Celebrities &       0.0 &         0.0 &  6 &     22 &   42 &  41 \\
\bottomrule
\end{tabular}
\end{table}

\begin{figure*}[t]
\centering
\captionsetup[subfigure]{labelformat=empty}
\subfloat{\includegraphics[width=0.19\textwidth]{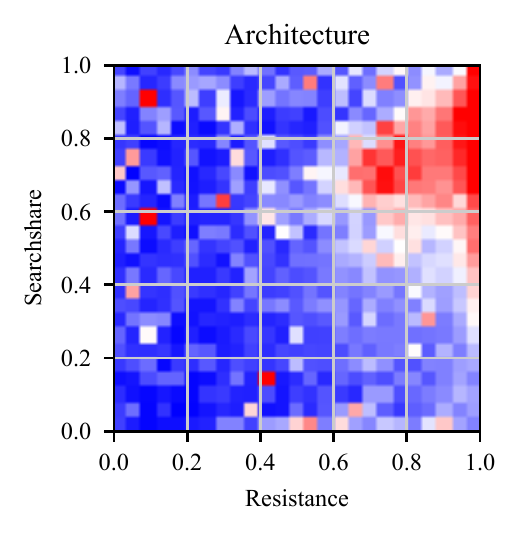}}
\subfloat{\includegraphics[width=0.19\textwidth]{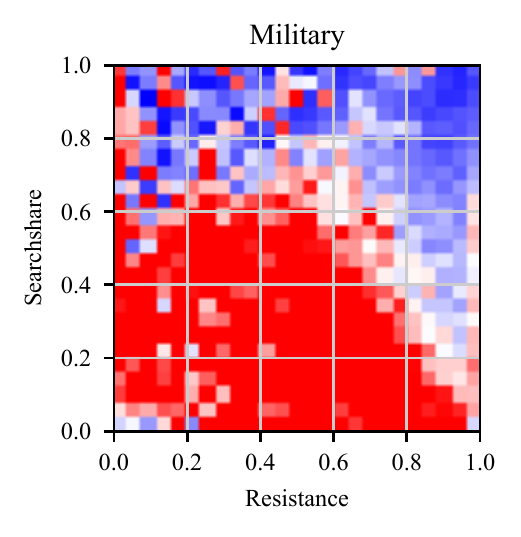}}
\subfloat{\includegraphics[width=0.19\textwidth]{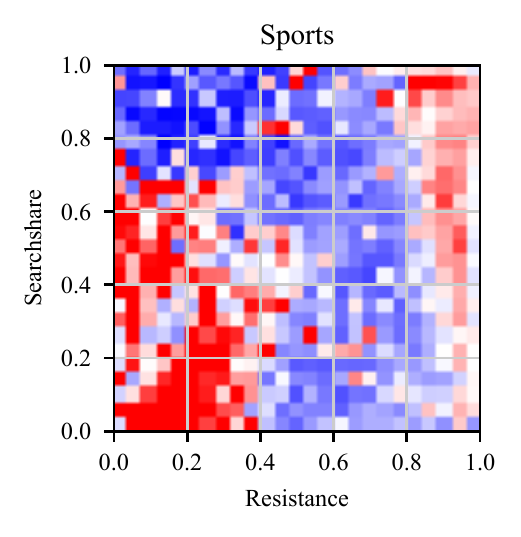}}
\subfloat{\includegraphics[width=0.19\textwidth]{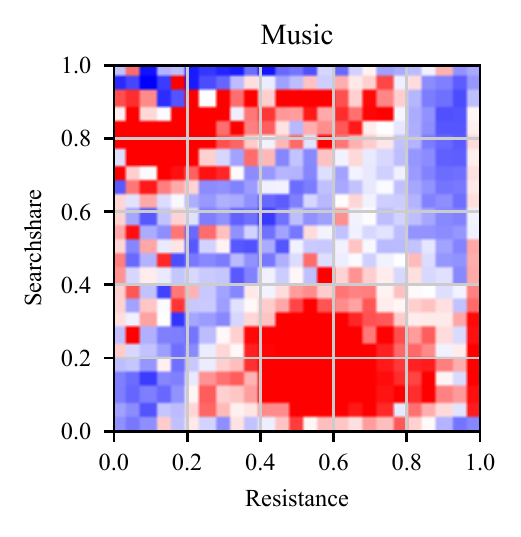}}
\subfloat{\includegraphics[scale=0.64]{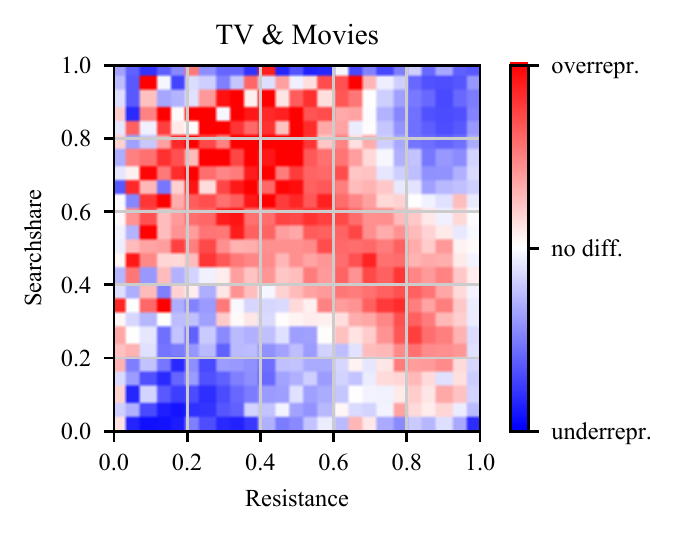}}
\vspace{-1em}
\caption{Relative difference of individual topics to the overall view distribution of searchshare vs. resistance (\cf Figure~\ref{fig:general_heatmap}(b)).
White denotes no relative difference, blue denotes underrepresentation (down to $0$), while red denotes overrepresentation (max. over all topics at $2$).
The figure highlights the differences between search-heavy and navigation-heavy topics compared to the all-articles baseline. ``Architecture'', exhibiting above-mean searchshare and resistance (\cf Figure \ref{fig:topic_views}) stands representative for six similarly distributed topics and mainly attracts search hits that it cannot pass on. ``Military'' shows an almost inverted pattern, mostly receiving as well as producing internal navigation. The bi-focal distribution of ``Sports'' can be found in ``Politics'' and ``Fine Arts \& Culture'' as well, while patterns for ``Music'' and ``TV \& Movies'' are more unique.}
\label{fig:topic_heatmaps}
\vspace{-1em}
\end{figure*}

\begin{figure*}[t]
\centering
\subfloat[k-core vs. searchshare]{\includegraphics[width=0.245\textwidth]{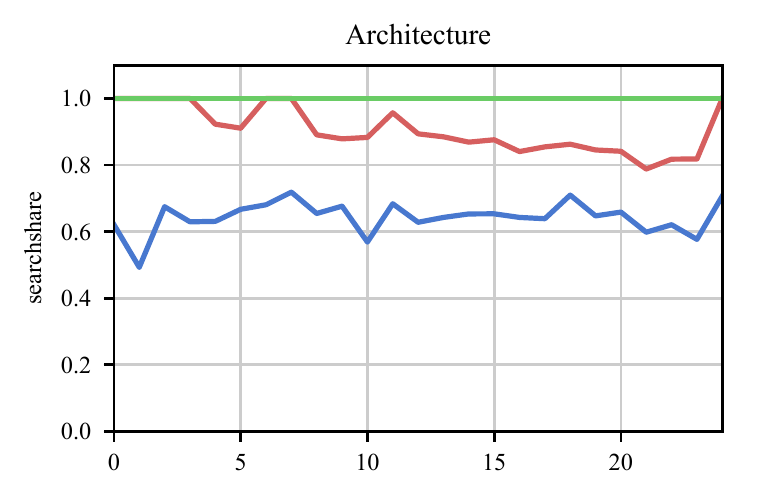}}
\subfloat[k-core vs. resistance]{\includegraphics[width=0.245\textwidth]{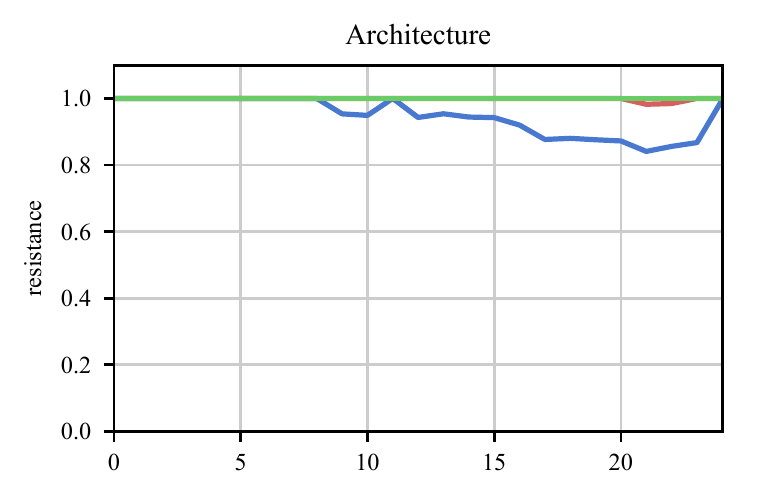}}
\subfloat[revisions vs. searchshare]{\includegraphics[width=0.245\textwidth]{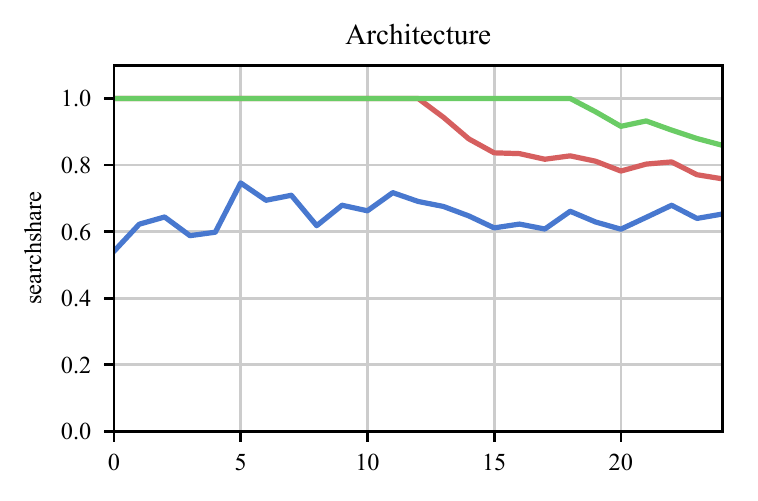}}
\subfloat[revisions vs. resistance]{\includegraphics[width=0.245\textwidth]{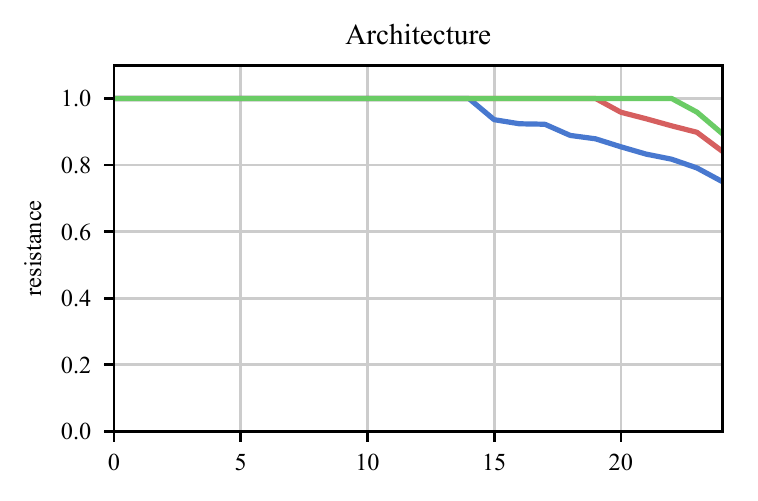}}

\subfloat[k-core vs. searchshare]{\includegraphics[width=0.245\textwidth]{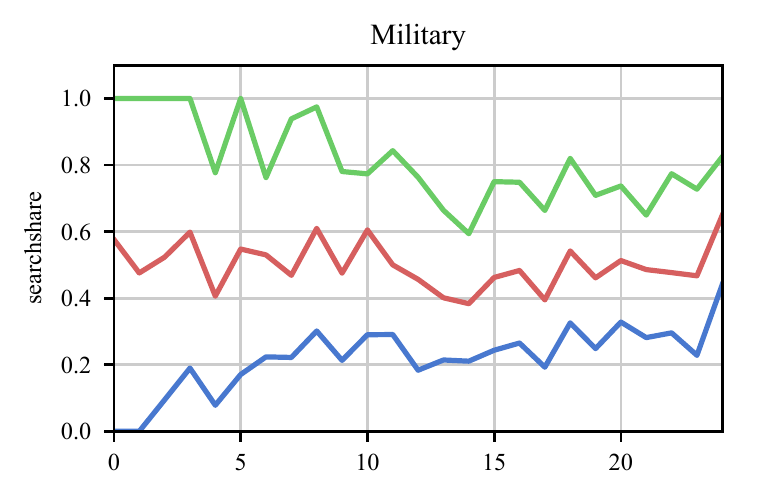}}
\subfloat[k-core vs. resistance]{\includegraphics[width=0.245\textwidth]{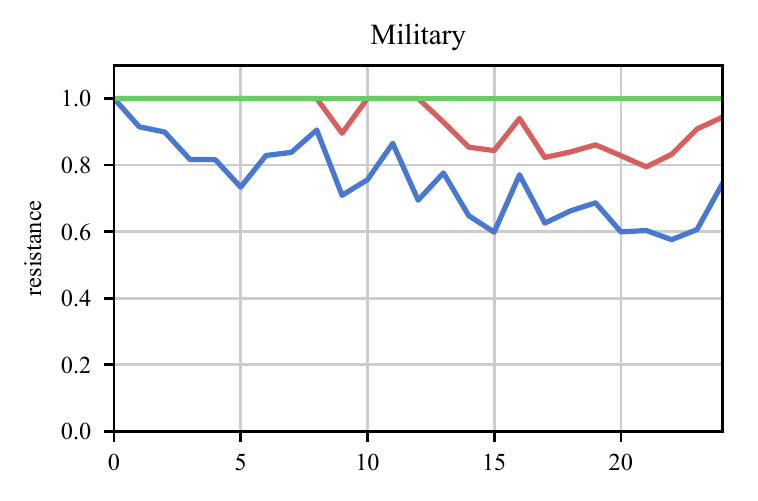}}
\subfloat[revisions vs. searchshare]{\includegraphics[width=0.245\textwidth]{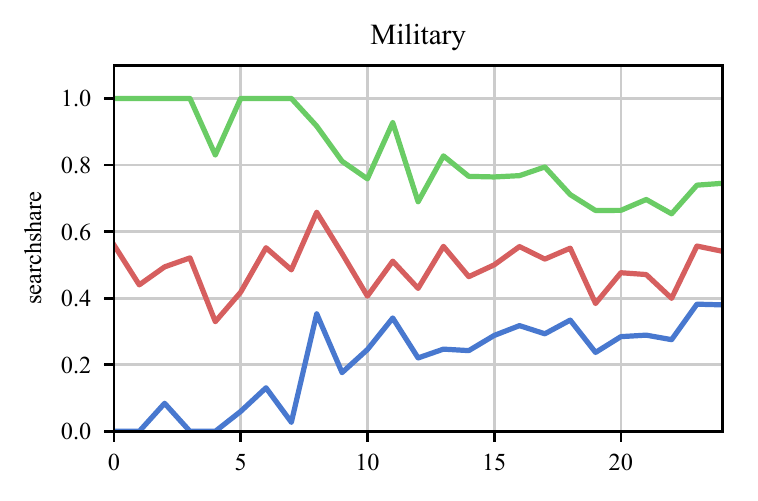}}
\subfloat[revisions vs. resistance]{\includegraphics[width=0.245\textwidth]{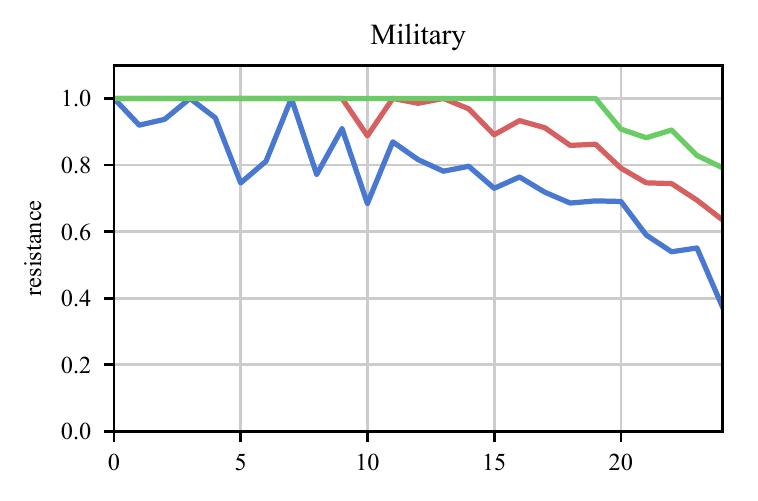}}
\vspace{-1em}
\caption{Relation of traffic features with network and content features. For a topic dominated by search (``Architecture'') and one dominated by navigation (``Military''), the figure shows the first (blue), second (red), and third (green) quartile of the article searchshare and resistance as function of its position in the network indicated by its k-core and editors' activity indicated by the number of revisions. Articles are divided into 25 bins by k-core and revision values. Apart from base-level differences of searchshare and resistance, the topics exhibit comparable trends, with the exception of searchshare not being influenced as much by network position or edit activity features for ``Military'' articles.%
}
\label{fig:topics_rev}
\vspace{-1em}
\end{figure*}

\subsection{Topic Features}
\label{sec:initail_topic}
Search-related popularity, navigability as well as other characteristics related to  traffic might be highly dependent on the topical domain of an article. We hence investigate the access behavior across Wikipedia's numerous article themes, represented by the 20 topics we have extracted.
Table \ref{tab:topics} provides descriptive statistics of these topics.
With 32\% ``TV and Movies'' is the topic with the most views while consisting of a mere 7.5\% of all articles on Wikipedia. ``Technology, Stubs''  and ``Architecture'' show an opposing dynamic, providing a large amount of articles, but relatively few views.\footnote{ ``Technology, Stubs'' is a compound of general Wikipedia:Stub articles and often short technology articles that were not sufficiently distinguishable by LDA. We exclude it from discussion here due to its ambivalent nature.} Overall, the amounts of articles and view counts are not strongly correlated. Consistent with previous research, we also observe that the popular articles are in general longer, relative old, and revised more often by more editors~\cite{spoerri2007popular}. 

A look at the distribution of searchshare and resistance in the overview provided by Figure~\ref{fig:topic_views} reveals the different access behaviors for Wikipedia topics. 
To examine these pronounced differences further, we set out to highlight the dissimilarities of the overall searchshare vs. resistance distribution for total views -- as shown in Figure \ref{fig:general_heatmap}\subref{fig:heatmap_pageviews_all} -- with the same distribution for the individual topics. To do so, we create heatmaps pinpointing the \textit{relative differences} of each topic to the baseline of the overall distribution. This is achieved by performing a bin-wise division of a topic's normalized view count for a given searchshare-resistance bin with the respective normalized bin for the general Wikipedia traffic behavior. %
The resulting heatmaps are shown in Figure~\ref{fig:topic_heatmaps}  for selected topics. They draw a clear picture of the over- and under-representation of certain article types (in terms of views) in each topic against the whole-system baseline. 
``Architecture'' in Figure~\ref{fig:topic_heatmaps} stands as one representative for a group of topics (``Biology'', ``Industry \& Chemistry'',  Research \& Education, ``Space \& Racing'') that all exhibit a very similar distribution with their article views occurring at high searchshare and high resistance, \ie, these topics are  mostly searched and not used for further navigation. In stark contrast, views for ``Military'' topics occur to the largest part in comparably low-resistance articles, that are mostly navigated to (views for ``UK \& Commonwealth'' are distributed almost analogously). 
``Sports'' reveals a similar inclination for \textit{nav-relay} types of articles attracting views, yet sports articles also frequently get accessed by search and abandoned immediately (closely related patterns: ``Fine Arts \& Culture'' and ``Politics''). Lastly, ``Music'' and ``TV \& Movies'' exhibit remarkably idiosyncratic distribution patterns, not mirrored by another topic. ``Music'' attracts many views in a \textit{search-relay} fashion, but on the other hand also explicitly acts as a ``dead end'' for internal navigation.

As ``maximally different'' topics in respect to these traffic patterns and with overall high view counts, we select ``Architecture'' for search-heavy topics, and ``Military'' for navigation-heavy topics to conduct a deeper analysis regarding article network, content and edit properties. While ``Architecture'' includes  articles covering popular buildings, landmarks and municipalities, ``Military'' consists of articles covering significant historic events often associated with violence such as wars and notable battles, along with many articles dedicated to military units, personnel and equipment (\cf~\cite{samoilenko2017History}).  For the general access behavior concerning the network, content and edit features, we again assign the articles to one of 25 bins according to their k-core and revision counts, compute the distribution of searchshare and resistance for each bin, and plot the quartiles (\cf Figure~\ref{fig:topics_rev}). 
For ``Architecture'', searchshare (a) initially decreases for increasing k-core but sees an uptick for very central nodes, and a very similar behavior can be observed for resistance (b). 
``Military'' is characterized by generally lower levels of both metrics, yet shares the trend of decreasing resistance with increasing k-core (e), meaning that for both topics, the more central articles in the network  are able to channel visitors into  Wikipedia, with the top-most central nodes excluded from this trend. 
Being edited more implies decreasing resistance for both topics ((d), (h)), although this trend reveals itself only for much higher revision counts for ``Architecture'', most likely to its generally higher resistance. 
Edit counts have no clearly distinguishable influence on ``Military'' articles' searchshare, for ``Architecture'' it, however, implies lower searchshare.

\para{Summary.}
The results presented in this section suggest that the content heavily influences the access behavior on Wikipedia. Particularly, topical domains are accessed differently, \ie, users prefer to access articles about architecture and landmarks mainly through search, whereas more historical articles about military actions are navigated.  Moreover, mature articles with high revision numbers and article located in the core of the network are more likely to channel traffic through Wikipedia, whereas articles located at the network periphery act as exit points.

\section{Modeling Access Behavior}
\label{sec:prediction_task}
Our previous analysis characterized the user access behavior on Wikipedia articles with respect to their traffic from search and navigation dependent on the article features. However, this analysis does not reveal the impact of the feature groups on the access behavior. To this end, we set out to model the access behavior on articles in order to measure the influence of each feature group. The higher the predicative performance of a feature group, the higher the influence of the group is on the role articles play with respect to the traffic (entry-exit and relay articles), and thus on the preferred information access form (search and navigation).

\begin{figure}[t]
\centering
\subfloat[searchshare]{\includegraphics[width=0.245\textwidth]{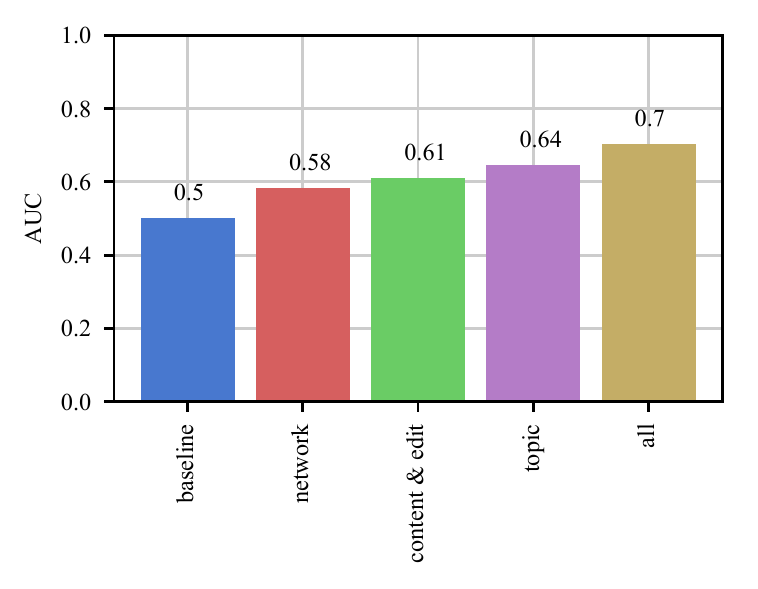}\label{fig:performance_searchshare}}
\subfloat[resistance]{\includegraphics[width=0.245\textwidth]{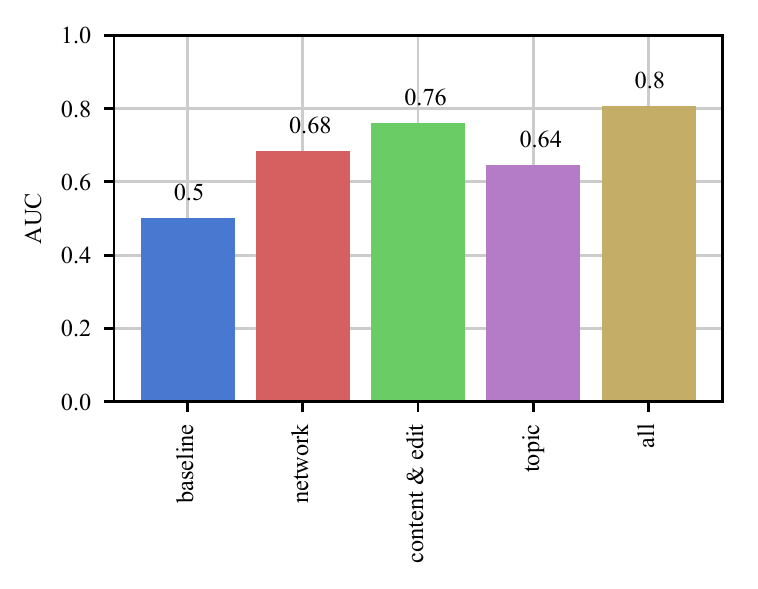}\label{fig:performance_resistance}}
\caption{Results. The figure shows the model performance (ROC AUC) for (a) searchshare and (b) resistance. Predicting searchshare is more challenging than predicting resistance. The article topic determines the preferred access behavior (search or navigation). However, position in the network, content maturity and presentation of the article are indicative for the resistance, and thus if an article will be an entry-exit point or a relay point for the traffic.}
\label{fig:performance}
\vspace{-1em}
\end{figure}

\para{Modeling searchshare.}
We ask, given a Wikipedia article, if it is possible to classify it as dominated by search, \ie, $searchshare>0.66$ or dominated by navigation, \ie,  $searchshare \leq 0.66$. The threshold used for the separation is the searchshare mean (\cf Section~\ref{sec:gab}). In our experiments, we consider four different sets of article features: (i) network features, \ie, in-, out-degree, k-core, (ii) content \&  edit features, \ie, number of revisions and editors, article size, number of tables, pictures, lists, and (iii) topic. For predicting the preferred access form, we fit a model using gradient boosting and evaluate the model's performance with ROC AUC. The model is trained using 10-fold cross validation at a balanced dataset. On this dataset random guessing results in 50\% accuracy, which is also used as baseline.
Figure~\ref{fig:performance}\subref{fig:performance_searchshare} shows the individual performance for each feature group, as well as the performance for the combination of all features. We observe that modeling searchshare is  difficult even with all features (AUC = 0.70). As expected, the network features are the least indicative (AUC = 0.58). Further, the topic feature predicts searchshare best (AUC = 0.64), which suggests strong user preferences for specific information access form, \ie, search or navigation for different topics. 

\para{Modeling resistance.}
To model the resistance of Wikipedia articles, \ie, they ability to relay traffic, we treat an article as a relay point if $resistance \leq 0.88$ and an exit point if $resistance>0.88$.  Again, the separation of the articles is based on the resistance mean (\cf Section~\ref{sec:gab}). We consider the same feature groups as for modeling searchshare and use random guessing as our baseline. For classifying the articles, we again utilize 10-fold cross validation to train a gradient boosting model on a balanced dataset. The performance is measured in terms of ROC AUC.
Figure~\ref{fig:performance}\subref{fig:performance_resistance} shows the individual classification performance for that task for each feature group. 
The content and edit features are the most important (AUC = 0.76). This makes the case for an influence of the way content is presented to the user on lowering or increasing the resistance of a page. 
Unlike for searchshare, the network features are indicative for the resistance of an article (AUC = 0.68). This suggests that the network position of an article influences the extent to which it channels traffic. The topic plays only a small role, which again highlights the importance of the quality of the content presentation and production process.

\para{Summary.}
In general, modeling article resistance is easier than modeling searchshare as suggested by the higher ROC AUC values. Modeling searchshare is challenging due to the influence of external events (\eg, the transition data exhibits high view numbers on articles about the Summer Olympics 2016), and the content diversity. However, investing in diverse content from different topics seems to be the best way for Wikipedia to attract people as the article's topic is the most indicative feature regarding searchshare. On the other hand, the content presentation, and the article's position in the network are decisive for its ability to relay traffic. 
\section{Related Work}
\label{sec:relwork}
Since the inception of the Web, researchers have been studying the user content consumption behavior. Initially, content has been accessed by traversing hyperlinks on the Web~\cite{kumar2010characterization}. This navigational user behavior on the Web and on Wikipedia is often modeled using well-established methods such as Markov chains~\cite{chierichetti2012web,singer2015hyptrails, singer2014detecting, page1999pagerank, pirolli1999} and decentralized search models~\cite{dimitrov2015role,helic2013models}. Numerous navigational hypotheses on Wikipedia have also been presented based on, \eg, click traces stemming from navigational games and on click data from server logs. For example, West and Leskovec observed a trade-off between similarity and popularity to the target article in the user sessions of Wikispeedia~\cite{west2012human}. Lamprecht \etal studied the general navigability of several Wikipedia language editions and showed how the Wikipedia article structure influences the user click behavior~\cite{lamprecht2016evaluating,lamprecht2017structure}. Dimitrov \etal conducted a large-scale study on the navigational behavior on Wikipedia. They found that users tend to select links located in the beginning of Wikipedia articles and links leading to articles located in the network periphery~\cite{dimitrov2017makes,dimitrov2016visual}. By constructing a navigational phase space from transition data, Gilderslave and Yasseri studied internal navigation on Wikipedia and identified articles with extreme, atypical, and mimetic behavior~\cite{gilderslave2017Inspiration}. 
Web content can be also discovered by formulating and executing a search query.
Kumar and Tomkins performed an initial characterization of the user search behavior~\cite{kumar2009characterization}, while Weber and Jaimes studied the search engine usage with respect to the users demographics, topics, and session length~\cite{weber2011uses}. 
Earlier Wikipedia reading behavior studies focused on explaining bursts, dynamics of topic popularity and search query analysis to Wikipedia~\cite{ten2012modeling,ratkiewicz2010characterizing,spoerri2007popular,waller2011search,lehmann2014reader}. A more recent study by Singer~\etal investigated the Wikipedia readers motivations~\cite{singer2017we}. By complementing a reader survey with server log data, they discovered specific behavior patterns for different motivations, \ie, bored readers tend to produce long article sequences spanning different topics. 
McMahon \etal focused on the interdependence between search engines, \ie, Google and Wikipedia~\cite{mcmahon2017substantial}. They showed that Google is responsible for generating high traffic to Wikipedia articles, although, in some cases traffic is reduced due to the direct inclusion of Wikipedia content in search results. Compared to our work, McMahon \etal concentrate on the peer production site and not on the content consumption. While there is a long line of research with respect to search and -- more so -- navigation, they have rarely been studied together which is the focus of this work.
\section{Discussion}
\label{sec:discussion}

As a general observation, our results shed light on the different roles of articles with respect to traffic entering and leaving Wikipedia. 
On one hand, an overwhelming amount of pages  attracts mostly direct search traffic and only little internal navigation, thanks to Wikipedia's strong symbiotic relationship with web search engines. 
Yet, notably, most of that traffic goes to articles that act mainly as exit points, \ie, users to not continue visiting Wikipedia directly afterwards. 
This is congruent with, but not necessarily because of, a pure ``look up'' nature of search.
Only a very small share of searched articles is responsible for relaying disproportionally large amounts of traffic into the rest of Wikipedia. 
We see that these articles are well-connected, more edited and more extensive than their exit counterparts, although we cannot yet conclude whether this is because of a ``worn path'' paradigm, wherein links and content are built because of the natural thematic positioning and suitability of an article to act as an entry point \textit{and} as a bridge to more content, or because the a-priori structure of these article facilitates the observed navigational patterns. 
A longitudinal study, which we plan for future work, could obtain more detailed insights on this co-evolution of structural features and navigation.
Furthermore, our data shows that articles, which are able to forward traffic, sit mostly at the very (k-)core of the link network.
However, this is not necessarily the case for being a receiver of navigation traffic, with searchshare values stabilizing already at lower k-cores -- and with inlinks not being more highly correlated with k-core than outlinks. This hints to the fact that -- to some extent -- users enter Wikipedia by search on more central articles, and then navigate outwards from more to less central nodes. This is consistent with previous findings studying navigation on Wikipedia~\cite{dimitrov2017makes}.

Regarding articles with different topical alignments, we see certain evidence that the thematic domain of a user's information pursuit seems  connected with the ``mode'' of how this information is attained. 
While the highly aggregate data used in this work does not allow for direct inferences as to the type of information retrieval in the continuum between a targeted and well-defined lookup and a completely serendipitous discovery process, we can nonetheless discern distinct patterns between article topics. Although ``Architecutre'' articles are not more devoid of in- or out-navigation opportunities than ``Military'' ones, they show far higher amounts of search views and exit points, while the latter one is navigated at a constantly high level, regardless of their connectedness. A possible explanation of the navigation-heavy behavior on ``Military'' articles is that people like follow paths through events in order to understand historical developments. 
%
%

%

%
%

%
For our analysis, we utilize publicly available clickstream data about Wikipedia. However, due to privacy restrictions, the data contains only (referrer, resource) pairs that occurred at least ten times during the data collection period. This could lead to a skewed view on the access behavior when contrasting search and navigation. 
For example, if an article is navigated in total much more than ten times over different links, but each individual link is transitioned less than ten times, all of these transitions will not be included in the data. In this case, the searchshare for this article might get substantially overestimated. Since this might occur specifically for articles with overall few page views, it may be a potential explanation for some findings, \eg, that article in the periphery of the link network show a stronger prevalence of search.

\section{Conclusion and Future Work}
\label{sec:conclusions}
In this work, we studied the prevalence of user access preferences across articles on Wikipedia. For that purpose, we introduced \emph{searchshare} and \emph{resistance} as two key features to characterize article traffic.
While we can identify search as the more dominant access paradigm compared to navigation on Wikipedia overall, we observe heterogeneous behavior at different types of articles. That is, depending on the article topic and other article properties, the share of navigation and search strongly varies, as well the amount of traffic an article relays to other Wikipedia pages. 
For example, articles on topics such as ``Military'' exhibit above average access by navigation, while topics such as ``Architecture'' show a strong prevalence of search.
Furthermore, edit activity on a an article and its position in the network is strongly correlated with its ability to relay traffic on Wikipedia.
Thus, we find overall that both, search and navigation play a crucial role for information seeking on Wikipedia.
In the future, we plan to extend our studies over time intervals and to other language editions in order to further explore cultural differences in the identified access patterns. 

\balance
\bibliographystyle{ACM-Reference-Format}
\bibliography{bib.bib} 

\end{document}